\documentclass[aps,prb,reprint,amsmath,amsfonts,amssymb,superscriptaddress]{revtex4-1}
\usepackage{graphicx}
\usepackage{dcolumn}
\usepackage{bm}
\usepackage{amsbsy}
\usepackage{braket}
\usepackage[pdfstartview=FitH,bookmarksnumbered=true,bookmarksopen=false]{hyperref}
\usepackage{xcolor}

\begin{document}

\title{Statistically related many-body localization in the one-dimensional anyon Hubbard model}

\author{Guo-Qing Zhang}
\affiliation{Guangdong Provincial Key Laboratory of Quantum Engineering and Quantum Materials, GPETR Center for Quantum Precision Measurement and SPTE, South China Normal University, Guangzhou 510006, China}
\affiliation{Frontier Research Institute for Physics,
South China Normal University, Guangzhou 510006, China}

\author{Dan-Wei Zhang} \email{danweizhang@m.scnu.edu.cn}
\affiliation{Guangdong Provincial Key Laboratory of Quantum Engineering and Quantum Materials, GPETR Center for Quantum Precision Measurement and SPTE, South China Normal University, Guangzhou 510006, China}
\affiliation{Frontier Research Institute for Physics,
South China Normal University, Guangzhou 510006, China}

\author{Zhi Li}
\affiliation{Guangdong Provincial Key Laboratory of Quantum Engineering and Quantum Materials, GPETR Center for Quantum Precision Measurement and SPTE, South China Normal University, Guangzhou 510006, China}

\author{Z. D. Wang}
\affiliation{Department of Physics and HKU-UCAS Joint Institute for Theoretical and Computational Physics at Hong Kong, The University of Hong Kong, Pokfulam Road, Hong Kong, China}
\affiliation{Frontier Research Institute for Physics,
South China Normal University, Guangzhou 510006, China}

\author{Shi-Liang Zhu}
\email[]{slzhu@nju.edu.cn}
\affiliation{Guangdong Provincial Key Laboratory of Quantum Engineering and Quantum Materials, GPETR Center for Quantum Precision Measurement and SPTE, South China Normal University, Guangzhou 510006, China}

\affiliation{Frontier Research Institute for Physics,
South China Normal University, Guangzhou 510006, China}

\date{\today}

\begin{abstract}

Many-body localization (MBL) has been widely investigated for both fermions and bosons, it is, however, much less explored for anyons. Here we numerically calculate several physical characteristics related to MBL of a one-dimensional disordered anyon-Hubbard model in both localized and delocalized regions. We figure out a logarithmically slow growth of the half-chain entanglement entropy and an area-law rather than volume-law obedience for the highly excited eigenstates in the MBL phase. The adjacent energy level gap-ratio parameter is calculated and is found to exhibit a Poisson-like probability distribution in the deep MBL phase. By studying a hybridization parameter, we reveal an intriguing effect that the statistics can induce localization-delocalization transition. Several physical quantities, such as the half-chain entanglement, the adjacent energy level gap-ratio parameter, {\color{black} the long-time limit of the particle imbalance}, and the critical disorder strength, are shown to be non-monotonically dependent on the anyon statistical angle. Furthermore, a feasible scheme based on the spectroscopy of energy levels is proposed for the experimental observation of these statistically related properties.

\end{abstract}


\maketitle

\section{Introduction}
Many-body localization (MBL) is the interacting analog of single-particle localization and extends the original work of Anderson~\cite{PhysRev.109.1492} with the effects of particle-particle interactions. There are two known classes of closed many-body systems: ergodic systems and MBL systems. Ergodic systems serve as a heat bath for themselves and thermalize after sufficient long unitary evolution, and, thus, the initial information of the systems is lost~\cite{PhysRevA.43.2046,rigol2008thermalization}. On the contrast, the emergence of local integral of motions caused by disorders, such as random potentials or interacting strengths leads to ergodicity breakdown and keeps the system in highly nonthermal states~\cite{PhysRevLett.44.1288,PhysRevLett.73.2607,PhysRevB.82.174411,PhysRevLett.118.196801}. The key ingredient for the many-body localization-delocalization transition is disorder via a mechanism similar to the Anderson localization. Various aspects of MBL systems are theoretically studied in the past few years with great progress, such as a criterion for many-body localization-delocalization phase transition proposed in Ref.~\cite{PhysRevX.5.041047}, high-energy eigenstates with power-law entanglement spectra in localized regions~\cite{PhysRevLett.117.160601}, and localization-induced real-complex transition in non-Hermitian MBL systems~\cite{PhysRevLett.123.090603}. MBL systems are robust against small perturbations and have the potential of storing initial state information for a long time and, hence, may be useful for dynamical quantum control and quantum memory devices. Current active experimental searches for MBL have been reported in ultracold atoms~\cite{Schreiber842,PhysRevLett.114.083002}, ultracold ions~\cite{smith2016many}, and superconducting circuits~\cite{Roushan1175,PhysRevLett.120.050507}.

Fractional statistics that interpolate boson statistics and fermion statistics was first proposed more than forty years ago in two-dimensional systems~\cite{Leinaas1977,PhysRevLett.48.1144,PhysRevLett.50.1395}. The particles that obey fractional statistics are anyons and the many-body wave function of the Abelian anyons acquires an additional phase $e^{i\theta}$ when exchanging two anyons on different sites where $\theta$ denotes the statistical angle. In the limit $\theta\rightarrow 0$, anyons become bosons whose wave functions remain invariant under particle exchange and when $\theta\rightarrow \pi$ anyons behave, such as fermions. Quasiparticles in the two-dimensional fractional quantum Hall effect obey fractional statistics and can be considered as anyons~\cite{PhysRevLett.48.1144,PhysRevB.41.12838}. Anyons play an important role as quasiparticles in topologically ordered states, and may be potentially useful in quantum information processing~\cite{KITAEV20032,RevModPhys.80.1083,Stern1179,00018732.2019.1594094}.

Fractional statistics was restricted in two-dimensional systems until Haldane introduced arbitrary dimensional fractional statistics~\cite{PhysRevLett.67.937}. Recently, a one-dimensional Hubbard model of fermions with the correlated hopping process has been proposed to realize fractional statistics~\cite{PhysRevLett.102.146404}. Alternative schemes for bosons with occupation-dependent hopping amplitudes by photon-assisted tunneling~\cite{keilmann2011statistically}, Raman-assisted hopping~\cite{PhysRevLett.115.053002} and lattice-shaking-induced resonant tunneling with potential tilts~\cite{PhysRevLett.117.205303} have also been proposed to realize anyons in one-dimensional optical lattices \cite{00018732.2019.1594094}. These proposals are based on the fractional Jordan-Wigner transformation by mapping anyons to bosons with a density-dependent tunneling parameter. Some exotic properties of one-dimensional anyons \cite{YJHao2012,WZhang2017,PhysRevLett.117.205303,PhysRevA.90.063618,JHHao2008} closely related to the statistical angle have been revealed, such as the statistically induced ground state phase transition~\cite{keilmann2011statistically,PhysRevA.94.013611,PhysRevB.97.115126,PhysRevB.99.165125}, the asymmetry of two-body correlations in the momentum space~\cite{PhysRevA.90.063618}, and the spatially asymmetric particle transport of interacting anyons~\cite{PhysRevLett.121.250404}. However, the MBL properties of anyons in disordered systems are largely unexplored.

In this paper, we numerically calculate several physical characteristics related to the MBL in a one-dimensional disordered (soft-core) anyon-Hubbard model in both localized and delocalized regions by using the numerical exact diagonalization (ED)~\cite{SLZhu2013,JMZhang2010,Weinberg2017,YLChen2020}. First, we present numerical evidence of the existence of the MBL phase in the anyon-Hubbard model. The half-chain entanglement entropy grows quickly in the ergodic phase and logarithmically slow in the localized region, respectively. The area-law growth of steady-state entanglement entropy for highly excited states is also explored. The calculated Poisson-like energy-level spacing statistics further indicates that the MBL phase exists in the anyon-Hubbard model with strong disorders, and the mean value of the gap-ratio parameter shows the $\theta$ dependence for various disorder strengths. We also find that the localization length for $\theta=\pi$ is larger than $\theta=0$. Then, by studying a hybridization parameter, we find that a localization-delocalization transition can be induced merely by the anyon statistic angle. Furthermore, several physical quantities, such as the half-chain entanglement, the adjacent energy-level gap-ratio parameter, the long-time limit of the imbalance, and the critical disorder strength, are found to be non monotonic functions as the statistical angle. Finally, we propose the scheme based on the spectroscopy of the energy-level techniques to observe the intrinsic properties of the MBL of anyons in a small system. In our scheme, both the mean value of the gap-ratio parameter and the inverse participation ratio can be extracted from the discrete-time Fourier transform of time-dependent two-point correlation functions.

The rest of this paper is organized as follows. In Sec.~\ref{sec:2}, we introduce the anyon-Hubbard model and its mapping to the Bose-Hubbard model with an occupation-dependent gauge field through the Jordan-Wigner transformation. Section~\ref{sec:3} is devoted to investigating the difference of ergodic and localized phases, studying the statistically induced localization-delocalization transition, and revealing the non monotonic dependence of critical disorder strength on the statistical angle. In Sec.~\ref{sec:4}, we propose the methods to experimentally observe the MBL in the system. A brief discussion and a short summary are presented in Sec.~\ref{sec:5}.

\section{\label{sec:2}Model and methods}

Let us first briefly introduce the anyon-Hubbard model and the fractional Jordan-Wigner transformation which exactly maps the anyon model to the boson model. The interacting anyon-Hubbard model
in the one-dimensional lattice reads~\cite{00018732.2019.1594094,keilmann2011statistically}
\begin{equation}
\hat{H}^a=-J\sum_{j=1}^{L-1}(\hat{a}_j^\dagger \hat{a}_{j+1}+\mathrm{H.c.})+\frac{U}{2}\sum_{j=1}^L\hat{n}_j(\hat{n}_j-1),
\end{equation}
where $J$ is the tunneling amplitude, $L$ is the lattice size, $U$ is the on-site interaction strength, and $\hat{n}_j=\hat{a}_j^\dagger \hat{a}_j$ is the anyon number operator on site $j$ with $\hat{a}_j^\dagger (\hat{a}_j)$ being the anyon creation (annihilation) operator on site $j$. This model in the clean case has been studied in Refs. \cite{00018732.2019.1594094,keilmann2011statistically}, and here, we consider disorders by adding random on-site potential $\sum_jh_j\hat{n}_j$ in $\hat{H}^a$, where $h_j\in[-W,W]$ and $W$ is the disorder strength. Anyons obey the generalized commutation relations,
\begin{align}\label{eq:gcr}
\hat{a}_{j} \hat{a}_{l}^{\dagger}-e^{-{i} \theta \operatorname{sgn}(j-l)} \hat{a}_{l}^{\dagger} \hat{a}_{j} &=\delta_{j l}, \\ \hat{a}_{j} \hat{a}_{l} &=e^{{i} \theta \operatorname{sgn}(j-l)} \hat{a}_{l} \hat{a}_{j},
\end{align}
where $\theta$ is the particle statistical angle and $\operatorname{sgn}$ is the sign operator with $\operatorname{sgn}(0)=0$. Thus, exchange particles between different sites will rise an additional phase factor $e^{i\theta}$ in the many-body wave function and particles on the same site behave the same as bosons. Anyons in the one-dimensional system can be mapped to bosons by the fractional Jordan-Winger transformation~\cite{keilmann2011statistically},
\begin{equation}\label{eq:jw}
\hat{a}_{j}=\hat{b}_{j} \exp \left({i \theta \sum_{l=1}^{j-1} \hat{n}_{l}}\right),\
\hat{a}_{j}^\dag=\exp \left({-i \theta \sum_{l=1}^{j-1} \hat{n}_{l}}\hat{b}_{j}^\dag\right ),
\end{equation}
where $\hat{b}_j (\hat{b}_j^\dag)$ is the boson annihilation (creation) operator. {\color{black}It is worth emphasizing that the particles are pseudo-fermionic in the $\theta=\pi$ limit and multiple particles can occupy the same site, thus, the on-site interaction is still relevant in this limit.} By making use of this anyon-boson mapping, the anyon-Hubbard model with on-site potential disorders can be rewritten under the boson operators
\begin{align}
\hat{H}^{b}=&-J \sum_{j}^{L-1}\left(\hat{b}_{j}^{\dagger} \hat{b}_{j+1} e^{i \theta \hat{n}_{j}}+\mathrm{H.c.}\right)\nonumber\\&+\frac{U}{2} \sum_{j}^{L} \hat{n}_{j}\left(\hat{n}_{j}-1\right)+\sum_jh_j\hat{n}_j. \label{eq:Boson_H}
\end{align}
Several schemes have been proposed to realize the Hamiltonian (\ref{eq:Boson_H}) in the absence of disorders with ultracold atoms in optical lattices \cite{00018732.2019.1594094,keilmann2011statistically,PhysRevLett.115.053002,PhysRevLett.117.205303}. Remarkably, the occupation-dependent synthetic gauge fields \cite{Lienhard2020,Goerg2019,Clark2018,Meinert2016} as the key ingredient and additional disordered potentials \cite{Schreiber842,PhysRevLett.114.083002} have been experimentally achieved.

Below, we implement an occupation-dependent tunneling scheme in the ED method to numerically handle the conditional-hopping Bose-Hubbard model. In the ED calculation, we use QuSpin~\cite{PAECKEL2019167998} with a modified Hamiltonian builder which inserts additional $e^{\pm in_j\theta}$ for all matrix elements of tunneling terms based on the occupation number $n_j$. The particle number can be directly read out from the constructed Fock state basis in the particle-conserving manifold. In the following numerical simulations, we set $J=1$ as the energy scale and use $U=1$ or $2$ in order to investigate the soft-core anyon cases. The open boundary condition is assumed in all of our numerical calculations.

 \begin{figure*}[tb]
\centerline{\includegraphics[width=0.95\textwidth]{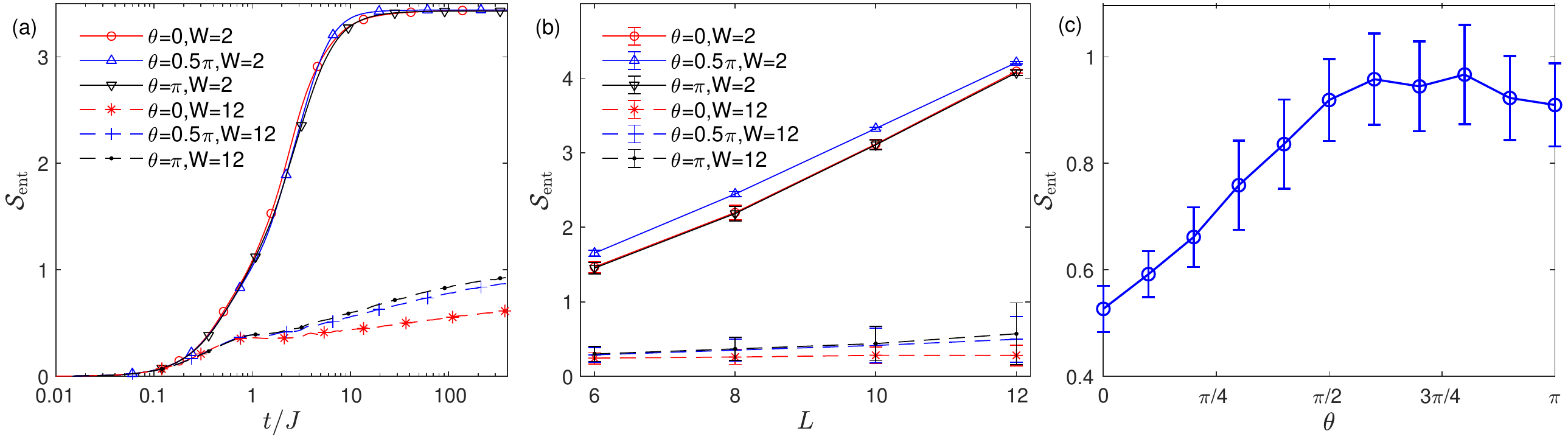}}%
 \caption{\label{fig:ent}(Color online) (a) The growth of half-chain entanglement entropy $\mathcal{S}_{\mathrm{ent}}(t)$ for the $L=10$ anyon-Hubbard model with two different disorder strengths $W=2, 12$ and three statistical angles $\theta=0, 0.5\pi, \pi$. (b) $\mathcal{S}_{\mathrm{ent}}$ of the highly excited eigenstates as a function of system size $L$. (c) $\mathcal{S}_{\mathrm{ent}}$ of the highly excited states as a function of $\theta$ for  $L=10$ and $W=8$. Other parameters are chosen as $J=1, U=2$, and all data is obtained by averaging over 20000, 10000, 2000, and 200 disorder realizations for $L=6,8,10, 12$ systems, respectively, in the half-filling manifold.}
 \end{figure*}

\section{\label{sec:3}Many-body localization}
In this section, we study the localization properties of the one-dimensional disordered anyon-Hubbard model and reveal their dependence of the anyon statistical angle $\theta$. The distinctions of entanglement growth, the dependence of half-chain entanglement on the system size, and the many-body energy level statistics indicate that both ergodic and localized phases exist in the anyon system. We show that the statistical angle $\theta$ has non-monotonic influence on the entanglement entropy, the mean value of the adjacent energy-level gap-ratio parameter, and the long-time evolution of the particle imbalance. These numerical results indicate the non monotonic dependence of the critical disorder strength on the statistical angle.

\subsection{Half-chain entanglement}

We first study the half-chain entanglement of many-body states in the anyon-Hubbard model. It was revealed that very weak interactions can significantly change the growth of entanglement in nonequilibrium many-body states driven by disordered Hamiltonians~\cite{PhysRevB.77.064426,PhysRevLett.109.017202,PhysRevLett.110.260601}. The entanglement entropy $\mathcal{S}_{\mathrm{ent}}$ can be defined as the von Neumann entropy,
\begin{equation}
\mathcal{S}_{\mathrm{ent}}=-\mathrm{Tr}\rho_A\ln\rho_A=-\mathrm{Tr}\rho_B\ln\rho_B,
\end{equation}
of the reduced density matrix of either side labeled by $A$ and $B$. $\mathcal{S}_{\mathrm{ent}}$ shows a characteristic logarithmically slow growth in the MBL phase and the saturate value is unbounded in the thermodynamic limit. Here, we consider a bipartition of equal half-chain $L_A=L_B=L/2$ and observe the logarithmic growth of $\mathcal{S}_{\mathrm{ent}}$ in our disordered anyon-Hubbard model. In our simulations, we implement the Chebyshev polynomials~\cite{mason2002chebyshev} to approximate the action of matrix exponential $\ket{\psi(t+\Delta t)}\approx e^{-i\hat{H}^b\Delta t}\ket{\psi(t)}$ at the time $t$, which can efficiently access the dynamical properties of soft-core anyons (at half-filling). The half-chain entanglement entropy is also calculated under this invariant subspace in order to avoid the diagonalization of the large reduced density matrix.

In Fig.~\ref{fig:ent} (a), we present the growth of half-chain entanglement entropy $\mathcal{S}_{\mathrm{ent}}(t)$ for $L=10$ anyon-Hubbard model in both ergodic and deep in the localized region for several statistical angles. At half-filling, we consider the initial state $\ket{\psi(0)}$ prepared in a product state where every even site is filled by an anyon. The time evolution of $\mathcal{S}_{\mathrm{ent}}(t)$ is obtained by averaging over 2000 disorder realizations with the results shown in Fig. ~\ref{fig:ent} (a). We can see that the entanglement entropies quickly increase from the initial time for both weak disorders (solid lines) and strong disorders (dashed lines), which correspond to the expansion of the wave package. The growth of the half-chain entanglement entropy cross from dephasing-dominated to transport-dominated dynamics~\cite{PhysRevLett.109.017202,PhysRevX.5.041047} and then increases logarithmically slow in time for all three simulated statistical angles for strong disorders, whereas it grows quickly and approaches the saturate value for weak disorders.

 \begin{figure*}[tb]
\centerline{\includegraphics[width=0.95\textwidth]{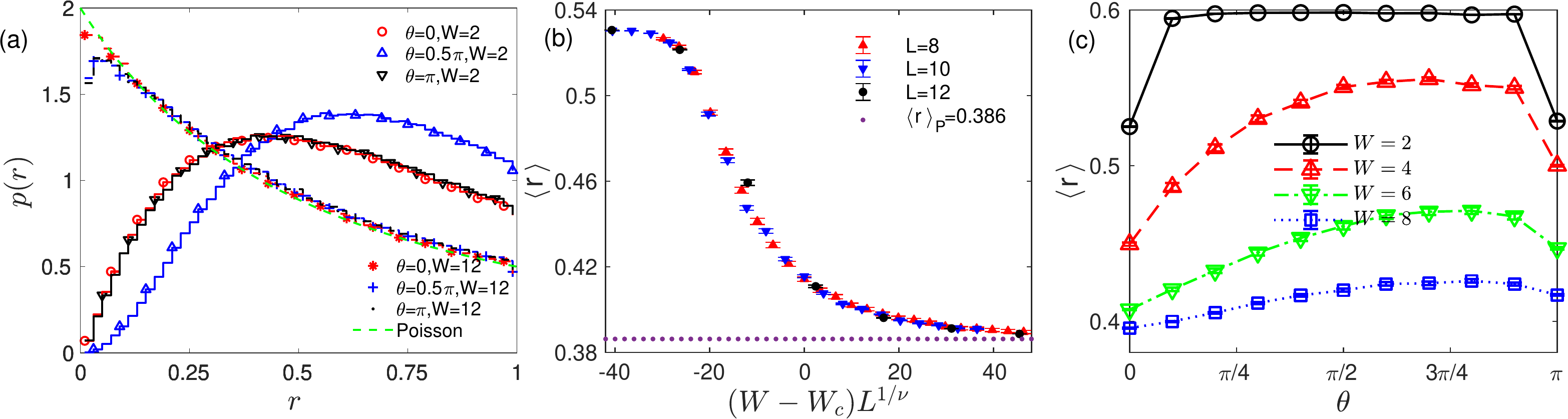}}%
 \caption{\label{fig:ls}(Color online) (a) The probability distribution of the gap-ratio parameter $p(r)$ for $L=10$ anyon-Hubbard model with two different disorder strengths $W=2, 12$ and three statistical angles $\theta=0, 0.5\pi, \pi$. The green dashed line is the Poisson distribution and is plotted as a guide to the eye. (b) The finite size scaling of mean value $\braket{r}$ with $\theta=0$, $\braket{r}$ collapse to a universal function $\braket{r(W,L)}=f[(W-W_c)L^{1/\nu}]$ for different system sizes $L=8, 10, 12$. (c) The mean value $\braket{r}$ as a function of statistical angle $\theta$ for different disorder strength $W$'s in $L=10$ systems. Other parameters are chosen as $J=1, U=2$, and all data is obtained by averaging over 20000, 10000, 2000, and 200 disorder realizations for $L=6, L=8, L=10$, and $L=12$ systems, respectively, in the half-filling manifold.}
 \end{figure*}

As the steady-state entanglement entropy $\mathcal{S}_{\mathrm{ent}}$ scales differently in the MBL and ergodic phases, we study its dependence on the system size for highly excited eigenstates in the two phases. By using the shift-invert spectral transformation $(\hat{H}^b-E_{\mathrm{shift}})^{-1}$ along with Krylov subspace methods~\cite{Hernandez:2005:SSF:1089014.1089019} with an energy shift $E_{\mathrm{shift}}$, we obtain those excited eigenstates nearest to $E_{\mathrm{shift}}=0$ up to $L=12$. We plot the averaged entanglement entropy $\mathcal{S}_{\mathrm{ent}}$ as a function of $L$ for two different disorder strengths $W=2,~12$ and three different statistical angles $\theta=0,0.5\pi,\pi$ in Fig.~\ref{fig:ent}(b). Here $\mathcal{S}_{\mathrm{ent}}$ is averaged over $20000$, $10000$, $2000$, $200$ disorder realizations for $L=6,~8,~10,~12$, respectively. In the ergodic phase for weak disorder ($W=2$, solid lines), the steady-state entanglement entropies of highly excited eigenstates increase significantly with system size for all three statistical angles. The reason for the non prefect linear dependence, here, lies in the fact that those eigenstates nearest to $E_{\mathrm{shift}}=0$ are not locating at the same position in the spectrum for different system size and disorder realizations. For strong disorder ($W=12$, dashed lines), the entanglement entropies show very weak dependence on system size. This phenomenon reveals the area-law entanglement in the deep MBL phase, which is different from the volume law in the ergodic phase~\cite{PhysRevLett.111.127201,PhysRevLett.115.187201}. The typical eigenstates of an ergodic system exhibit thermal volume-law entanglement according to the eigenstate thermalization hypothesis, and this volume law will be broken down by strong enough disorders, and the entanglement entropy scales with the area between two bipartite subsystems $A$ and $B$, which means $\mathcal{S}_{\mathrm{ent}}$ is approximately independent of the system size for one-dimensional systems. These MBL eigenstates are short-range entangled and locally correlated near the boundary of two subsystems~\cite{PhysRevLett.111.127201}.

We further calculate $\mathcal{S}_{\mathrm{ent}}$ of highly excited states as a function of the statistical angle $\theta$, with the results shown in Fig.~\ref{fig:ent} (c). Here $\mathcal{S}_{\mathrm{ent}}$ is calculated from states nearest to $E_\mathrm{shift}=0$ of the $L=10$ anyon-Hubbard systems and is averaged over $2000$ disorder realizations with $W=8$ [other parameters are the same with those in Fig.~\ref{fig:ent} (b)]. The half-chain entanglement entropy shows a non monotonic relation with $\theta$, which first grows, then, decreases when increasing $\theta$ with $\mathcal{S}_{\mathrm{ent}}(\theta=\pi)$ larger than $\mathcal{S}_{\mathrm{ent}}(\theta=0)$. Note that $\mathcal{S}_{\mathrm{ent}}$ changes from volume law to area law when the eigenstate is localized and can reflect the localization property in some aspects.

\subsection{Energy-level statistics}

The adjacent energy levels of a many-body Hamiltonian show different spectral statistics in the localized and ergodic phases. In the ergodic phase, the energy levels of large amounts of disorder realizations are described by random matrix theory, particularly, by the Gaussian orthogonal ensemble (GOE) for real symmetric matrices and Gaussian unitary ensemble (GUE) for complex Hermitian matrices~\cite{PhysRevA.42.1027,livan2018introduction,haake1991quantum}. In the MBL phase, nearby eigenstates that localized in the Fock space without level repulsion do not interact with each other and the nearest energy levels show Poisson statistics~\cite{PhysRevB.75.155111}. For those ED solvable finite-size systems, energy levels usually vary smoothly between GOE/GUE and Poisson statistics when increasing the disorder strength $W$. In order to avoid energy unfolding, a dimensionless gap-ratio parameter can be used to characterize statistics between adjacent energy-level gaps~\cite{edelman2005random,PhysRevB.75.155111,PhysRevLett.122.040606}. The gap-ratio parameter is defined as~\cite{PhysRevB.75.155111,PhysRevLett.122.040606}
\begin{equation}
r_n=\frac{\mathrm{min}\{\delta_n,\delta_{n-1}\}}{\mathrm{max}\{\delta_n,\delta_{n-1}\}},
\end{equation}
where $\delta_n=E_{n+1}-E_n$ is the adjacent energy level gap. The Poisson distribution of $r$ is $p(r)=2/(1+r)^2$ and has the mean value $\braket{r}_P=2\ln2-1$.

We numerically calculate the gap-ratio parameter of the anyon-Hubbard model for weak and strong disorder strengths and three different statistical angles with the results shown in Fig. \ref{fig:ls}. In Fig.~\ref{fig:ls} (a), the probability distribution of gap-ratio parameter $p(r)$ in the MBL phase ($W=12$, dashed lines) shows Poisson like behavior whereas $r$ in the ergodic phase ($W=2$, solid lines) has probability distribution of GOE ($\theta=0,\pi$) or GUE ($\theta=0.5\pi$). The green dashed curve is exactly the Poisson distribution plotted as a guide to the eye. Here, the system size is limited to $L=10$ whose Hilbert space is $2002$ in the half-filling manifold, and 2000 disorder realizations are averaged. For different statistical angles, the probability distribution $p(r)$ behaves similarly for strong disorders but distinguishable in the ergodic phase. For the weak disorder strength, anyons with statistical angle $\theta=0$ or $\pi$ are more likely to be localized than $\theta=0.5\pi$.

To further investigate the energy-level statistics, we analyze the relationship among the mean value of gap-ratio parameter $\braket{r}$, system size $L$, and disorder strength $W$. It is revealed in other many-body systems that the mean value $\braket{r}$ is a universal function of $(W-W_c)L^{1/\nu}$~\cite{vanNieuwenburg9269,xu2019stochastic}, where $W_c$ is the critical value of ergodic-MBL transition and $\nu$ is a critical exponent. We fit this universal function in Fig.~\ref{fig:ls} (b) for three different system sizes, $L=8, 10, 12$, averaged from $10000$, $2000$, $200$ disorder realizations, respectively, with the statistical angle $\theta=0$. By choosing $W_c\approx 5.5$ and $\nu\approx 1.1$, we can see that these three curves approximately collapse to the same curve which stands for a universal function $\braket{r(W,L)}=f[(W-W_c)L^{1/\nu}]$. The dotted line indicates the Poisson limit $\braket{r}_P\approx 0.386$, and it is clear that the mean value $\braket{r}$ tends to this limit when increasing disorder strength $W$. We also find similar behaviors for other statistical angles, but the corresponding critical disorder strength $W_c$ is quantitatively different. Due to the small system size and limited disorder realizations available in the ED method, we are unable to figure out the difference of critical exponent $\nu$'s for different statistical angle $\theta$'s.

Furthermore, we numerically obtain the mean value $\braket{r}$ as a non-monotonic function of $\theta$, which is depicted in Fig.~\ref{fig:ls} (c) for four different disorder strengths, with the parameters $J=1, U=2$, and $L=10$ in the half-filling manifold. For moderate and strong disorder strengths ($W=4,6,8$ for the red upward triangle, the green downward triangle, and the blue square, respectively), the mean values $\braket{r}$ show a non-monotonic relation with $\theta$, similar to those observed from Figs. \ref{fig:ent} (c) and \ref{fig:wc} (c). It is also obvious that the values of $\braket{r}$ are larger for $\theta=\pi$ than those for $\theta=0$, which implies that the fermions (anyons with $\theta=\pi$ at the anyon-Hubbard model) is more difficult than bosons (anyons with $\theta=0$) to be localized, which is consistent with the results shown in Fig.~\ref{fig:ent} (c). For weak disorder $W=2$ (black circle), the mean value $\braket{r}$ forms a plateau with $\braket{r}\approx0.53$ at $\theta=0$ and $\theta=\pi$. For $\theta$'s in between $0$ and $\pi$, the Hamiltonian becomes complex Hermitian, and $\braket{r}\approx0.6$ .

{\color{black} We present a heuristic argument to understand the $\theta$-dependent localization features. In the simplest case with a $2\times2$ Hermitian matrix with the eigen-energies $E_{1,2}$, the required numbers of parameters for the vanishing energy level spacing $|E_2-E_1|$ are different for real and complex matrices. A general real symmetric $2\times2$ matrix can be written as \[H_\mathrm{r}=\begin{pmatrix}
a & c\\ c & b
\end{pmatrix},\] where the matrix element $a$ can be taken as the energy unit and $b$ and $c$ can be tuned to vanish the energy-level spacing. Thus, two out of three parameters (elements $b$ and $c$) in a real symmetric matrix should be controllable for a vanishing energy-level spacing. However, for a general complex Hermitian matrix \[H_\mathrm{c}=\begin{pmatrix}
a & c_1+ic_2\\ c_1-ic_2 & b
\end{pmatrix},\] the off-diagonal element has an imaginary part, and three out of four parameters ($b, c_1$ and $c_2$) should be controllable. The level crossing resistance for complex Hermitian Hamiltonians are greater than real symmetric ones, and the mean value $\braket{r}$ of the complex Hamiltonians is generally larger than real ones (see Ref.~\cite{haake1991quantum} for a detailed interpolation). For the one-dimensional anyon-Hubbard model, the statistical angle $\theta$ in between $\theta=0$ and $\theta=\pi$ leads the off-diagonal elements of the Hamiltonian to complex and show a larger mean value $\braket{r}$.}

\subsection{Quench dynamics}

\begin{figure}[tb]
\centerline{\includegraphics[width=0.45\textwidth]{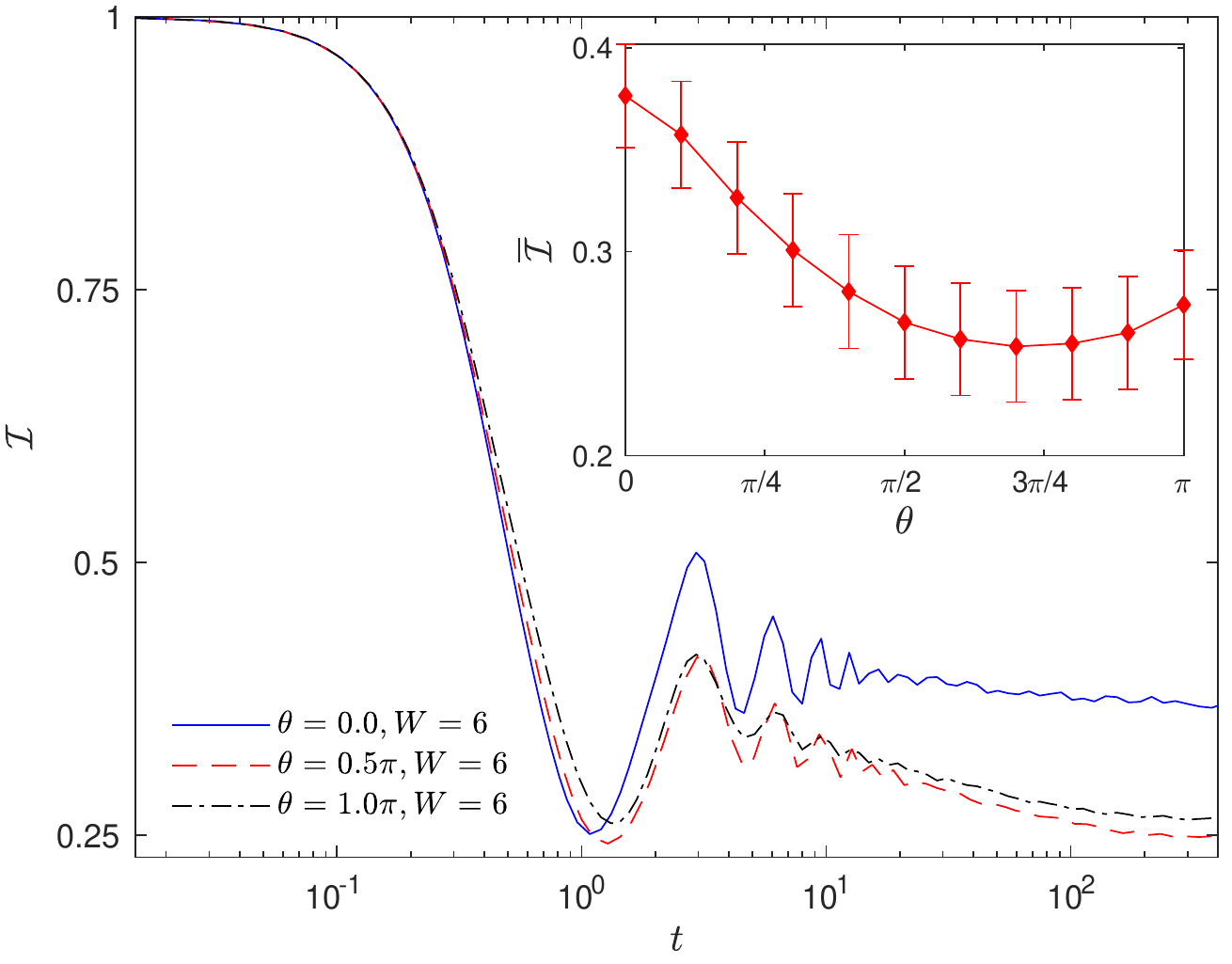}}%
 \caption{\label{fig:imb}(Color online) {\color{black} The time evolution of the particle imbalance $\mathcal{I}$ for three statistical angles $\theta=0,0.5\pi,\pi$ in the half-filling manifold. The inset plot is the long-time limit imbalance $\overline{\mathcal{I}}$ as a function of $\theta$. Parameters are chosen as $J=1, U=2, L=10$, and $W=6$, and the results are averaged over $2000$ disorder realizations.}}
 \end{figure}

{\color{black} Localization of particles in a quench dynamics can provide addition evidence of ergodicity breakdown and is experimentally observable in disordered systems~\cite{PhysRevLett.100.013906,PhysRevLett.101.255702}. In this subsection, we numerically study the quench dynamics and reveal the non-monotonic $\theta$-dependence of the long-time limit imbalance [see Eq.\ref{avgimb}]. The even-odd particle imbalance is defined as ~\cite{vanNieuwenburg9269}
\begin{equation}
\mathcal{I}(t)=\frac{\hat{n}_e(t)-\hat{n}_o(t)}{\hat{n}_e(t)+\hat{n}_o(t)},
\end{equation}
where $\hat{n}_e(t)=\sum\hat{n}_{2i}(t)$ and $\hat{n}_o(t)=\sum\hat{n}_{2i+1}(t)$ are the sum of anyons on each even and odd site at time $t$, respectively. The initial state is prepared in an out-of-equilibrium density configuration (i.e., an anyon on each even site). The system will keep a non-vanishing imbalance even after a long time evolution due to the breakdown of the ergodicity.

Figure~\ref{fig:imb} shows the time evolution of the imbalance $\mathcal{I}$ for three statistical angles $\theta=0,0.5\pi,\pi$ in the half-filling manifold, with $J=1, U=2, L=10, W=6$. The even-odd imbalance $\mathcal{I}$ has a finite value for a long time non-equilibrium evolution for all three statistical angle $\theta$'s. The inset plot of Fig.~\ref{fig:imb} shows the long-time limit of the imbalance,
\begin{equation}\label{avgimb}
\overline{\mathcal{I}}=\frac{1}{t_{f_2}-t_{f_1}}\int_{t_{f_1}}^{t_{f_2}}\mathcal{I}(t)dt,
\end{equation}
which is averaged from $t_{f_1}=200$ to $t_{f_2}=210$ as a function of statistical angle $\theta$. This long-time limit imbalance, consisting with half-chain entanglement and mean value $\braket{r}$, shows a $\theta$ dependent non-monotonicity. The largest value of $\overline{\mathcal{I}}$ suggests anyons with $\theta=0$ are more localized than any other $\theta$'s under the same disorder strength.}

\subsection{Localization length}

\begin{figure}[tb]
\centerline{\includegraphics[width=0.48\textwidth]{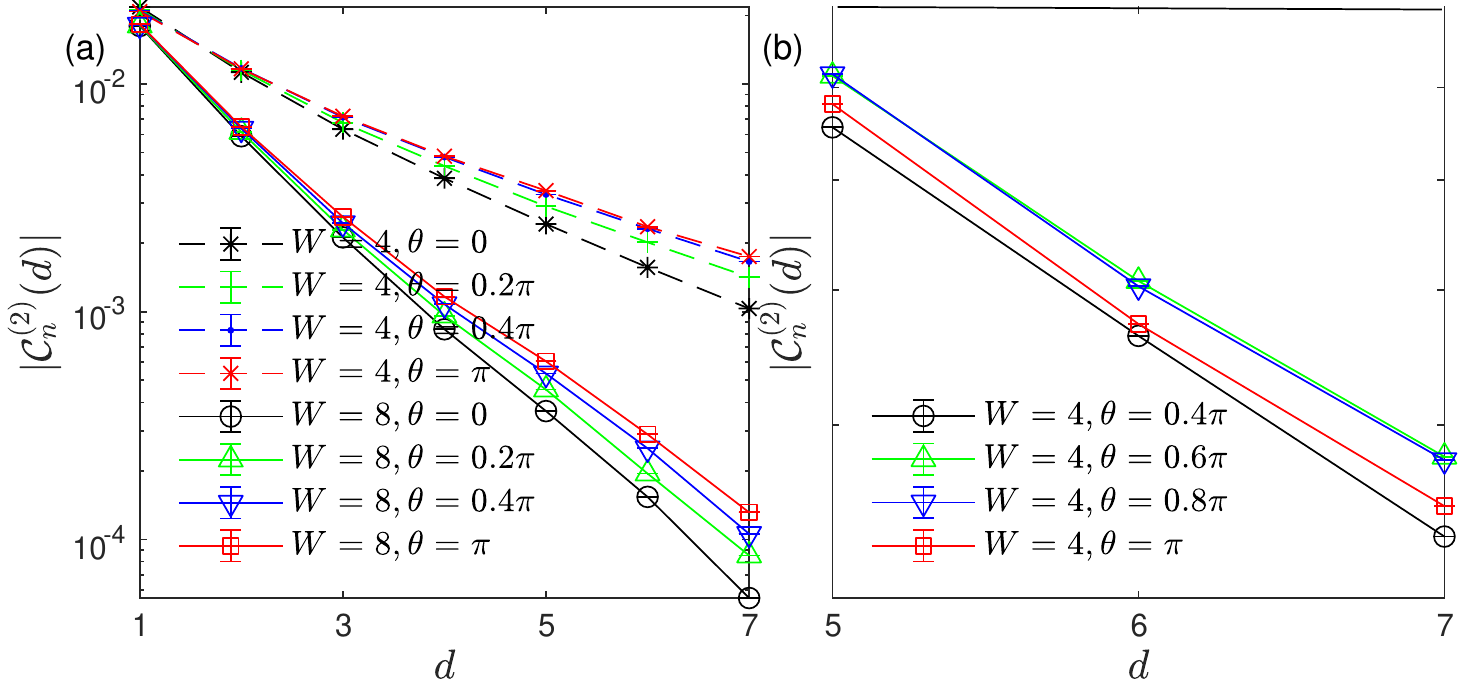}}%
 \caption{\label{fig:ll}(Color online) The two-particle correlation ${\mathcal{C}_n^{(2)}(d)}$ as a function of the distance $d$ for different statistical angle $\theta$ and disorder strength $W$, with the parameters  $J=1, U=2$, and $L=14$ in the two-particle manifold. All data are averaged over $10000$ disorder realizations.}
 \end{figure}

The localization length is one of the standard measures of localization in single-particle systems. When considering interactions, it is difficult to derive the localization length exactly. The interacting localization length can be extracted from two-particle correlations, which is given by~\cite{PhysRevLett.105.163905}
\begin{equation}
\mathcal{C}_n^{(2)}(i,j)=\bra{\psi_n}a^\dag_ia^\dag_ja_ja_i\ket{\psi_n},
\end{equation}
where $\ket{\psi_n}$ is the $n$-th eigenstate of the many-body Hamiltonian. Then the distance dependent
\begin{equation}
\mathcal{C}_n^{(2)}(d)=\sum_i\mathcal{C}_n^{(2)}(i,i+d)/(L-d)
\end{equation}
is the average of two-particle correlation with the same distance $j-i=d$. Near the localized phase, this two-particle correlation falls off exponentially with the distance,
\begin{equation}
\mathcal{C}_n^{(2)}(d)\sim e^{-(d/\xi)},
\end{equation}
where $\xi$ is defined as the localization length in the interacting systems~\cite{PhysRevB.82.174411}. We calculate  ${\mathcal{C}_n^{(2)}(d)}$ in the two-particle manifold with the parameters $J=1, U=2, L=14$, and $\ket{\psi_n}$ chosen to be the eigenstate at the middle of the energy spectrum. It is clear from Fig.~\ref{fig:ll}(a) that ${\mathcal{C}_n^{(2)}(d)}$ approximately falls off exponentially when increasing the distance $d$. For the same disorder strength (solid lines for $W=8$ and dashed lines for $W=4$), the decay rate of ${\mathcal{C}_n^{(2)}(d)}$ is the largest and smallest for $\theta=0$ and $\theta=\pi$, respectively. This means the localization length $\xi(\theta=\pi)>\xi(\theta=0.4\pi)>\xi(\theta=0.2\pi)>\xi(\theta=0)$, and the anyon-Hubbard model with $\theta=\pi$ is harder to localize than $\theta=0$. {\color{black} To see the relation between $\xi$ and $\theta$ more clearly, we plot additional curves of ${\mathcal{C}_n^{(2)}(d)}$ for $\theta=0.6\pi$ and $0.8\pi$ alone with $\theta=0.4\pi$ and $\pi$ in Fig.~\ref{fig:ll}(b). The decay rate of the two-particle correlation indicates that the localization length $\xi(\theta=0.8\pi)\approx\xi(\theta=0.6\pi)>\xi(\theta=\pi)>\xi(\theta=0.4\pi)$, which reveals the non-monotonic dependence of localization lengths on the statistical angle.}

\subsection{Statistically induced localization-delocalization transition}

 \begin{figure*}[tb]
\centerline{\includegraphics[width=0.9\textwidth]{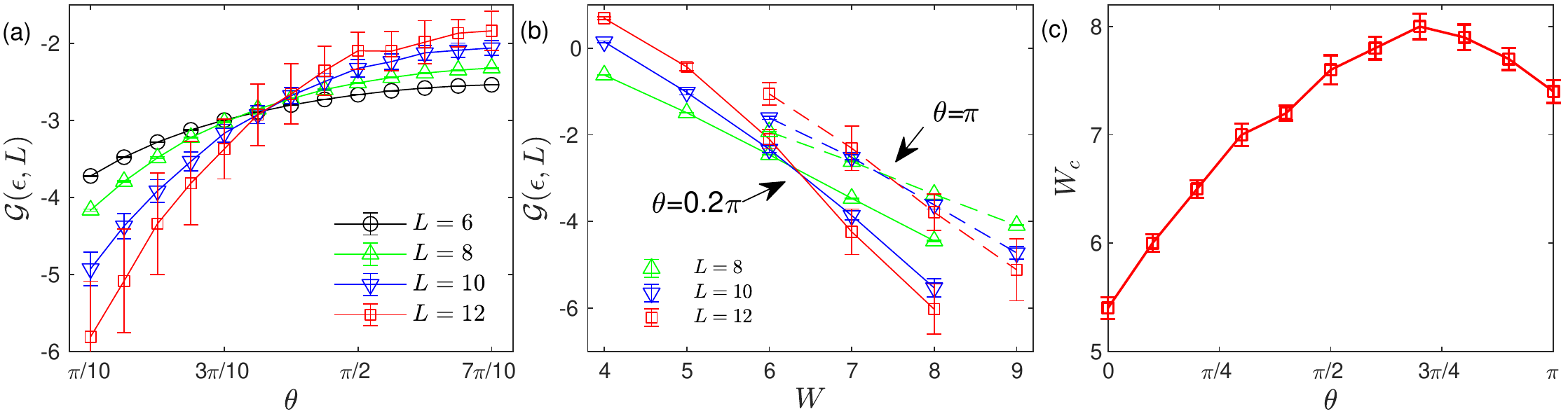}}%
 \caption{\label{fig:wc}(Color online) (a)  The hybridization parameter $\mathcal{G}(\epsilon,L)$ as a function of statistical angle $\theta$ for different system sizes $L=6,8,10,12$ {\color{black}with the disorder strength $W=7$}. (b) The hybridization parameter as a function of disorder strength $W$ for different system sizes {\color{black}$L=8,10,12$ and $\theta=0.2\pi$ (three solid lines), $\theta=\pi$ (three dashed lines).} (c) The critical disorder strength $W_c$ determined by  the crossover of hybridization parameter $\mathcal{G}(\epsilon,L)$ for different system sizes as a function of $\theta$. Some $20000, 10000, 2000, 200$ disorder realizations are used for $L=6, L=8, L=10$, and $L=12$ respectively in the half-filling manifold. Other parameters $J=1, U=2$.}
 \end{figure*}

Up to now, we have shown the existence of the MBL phase in the one-dimensional anyon-Hubbard model, and find that the physical quantities non-monotonically depends on the statistical angle $\theta$. In this subsection, we uncover an intriguing phenomenon that the anyonic statistics may induce the localization-delocalization transition at a fixing disorder strength $W$.

In order to detect the localization-delocalization transition in many-body interacting systems, we adopt the hybridization parameter $\mathcal{G}(\epsilon,L)$ introduced in Ref.~\cite{PhysRevX.5.041047}. The hybridization parameter is given by
\begin{equation}
\mathcal{G}(\epsilon,L)=\ln \frac{|\braket{\psi_{n+1}|\hat{V}|\psi_n}|}{E_{n+1}^\prime-E_n^\prime},
\end{equation}
where $\epsilon=(E_n-E_{\mathrm{min}})/(E_{\mathrm{max}}-E_{\mathrm{min}})$ is the energy density with $E_n$ in ascending order, $E_n^\prime=E_n+\braket{\psi_n|\hat{V}|\psi_n}$ is the modified energy, $\ket{\psi_n}$ is the eigenstate corresponding to energy $E_n$, and $E_{\mathrm{max}} (E_{\mathrm{min}}$) is the highest excited (ground) energy, and $\hat{V}$ is a perturbation operator. This parameter characterizes the hybridization of nearest eigenstates induced by the perturbation. Typically, $\mathcal{G}(\epsilon,L) \propto -\kappa L$, and $\kappa=0$ separates localized states ($\kappa > 0$) from delocalized states ($\kappa < 0$)~\cite{PhysRevX.5.041047}. Thus $d\mathcal{G}(\epsilon,L)/dL$ can be used to detect the localization-delocalization transition. Here we choose $\hat{V}=\hat{a}^\dagger_{L/2}\hat{a}_{L/2+1}$ as the perturbation operator and calculate $\mathcal{G}(\epsilon,L)$ as a function of statistical angle $\theta$ for different system size $L$'s, and set $\epsilon$ to where the delocalized phase is most robust in the whole energy spectrum. {\color{black}In numerical calculations, we set the value of $\epsilon$ and obtain the crossover of $\mathcal{G}(\epsilon,L)$ for different $L$'s, the critical disorder value is chosen to be the maximum crossover value when turning $\epsilon$ from 0 to 1. We note that choosing other perturbation operators would lead to similar results \cite{PhysRevX.5.041047}.}

In Fig.~\ref{fig:wc} (a), we plot $\mathcal{G}(\epsilon,L)$ for the half-filling anyon-Hubbard model with $J=1, U=2, W=7$ and four different $L$'s ($L=6$, black circle; $L=8$, upward triangle; $L=10$, blue downward triangle; $L=12$, red square). Some $20000, 10000, 2000, 200$ disorder realizations are averaged for $L=6,8,10,12$, respectively. The hybridization parameter $\mathcal{G}(\epsilon,L)$ decreases (increases) when enlarging system size $L$ for small (large) $\theta$, and the crossover indicates the localization-delocalization transition where $\kappa=0$. It is clear from Fig.~\ref{fig:wc} (a) that there is a transition from localized to delocalized phase transition at $\theta\approx 0.4\pi$ when disorder strength is fixed at $W=7$. To further clarify the different critical disorder strength $W_c$'s for different statistical angle $\theta$'s, we calculate $\mathcal{G}(\epsilon,L)$ as a function of disorder strength $W$ with the parameters $J=1, U=2$ and the results are plotted in Fig.~\ref{fig:wc} (b). The crossover for different system size happens at $W\approx 6.5$ and $W\approx 7.4$ for $\theta=0.2\pi$ and $\theta=\pi$, respectively. From this aspect, the critical disorder strength $W_c$ also shows $\theta$ dependence.

It is found that the critical disorder strength $W_c$ also has a non-monotonic relationship with the statistical angle $\theta$. First, we examine the critical disorder strength $W_c$ as a function of statistical angle $\theta$ by using the hybridization parameter $\mathcal{G}(\epsilon,L)$. As shown in Fig.~\ref{fig:wc} (c) [with the same parameters as Fig.~\ref{fig:wc} (b)], the critical disorder $W_c$ first increases with anyonic statistics $\theta$ and, then, decreases when $\theta\gtrapprox 0.7\pi$. The critical value for $\theta=0$ is $W_c\approx5.4$, whereas the value for $\theta=\pi$ is much larger ($W_c\approx7.3$). Such a non monotonic relation for the critical localization-delocalization transition value of $W_c(\theta)$ is similar as from other quantities related to the ergodic-MBL transition [see Figs.~\ref{fig:ent} (c), \ref{fig:ls} (c), and \ref{fig:imb}].

\section{\label{sec:4} Proposal of Experimental observations}

A recently developed many-body spectroscopy technique is able to resolve the energy levels of the interacting system~\cite{Roushan1175} and makes it possible to observe the properties of the MBL in a realistic experiment setup. We simulate the spectroscopy of energy levels in $L=9$ sites systems with a maximum of two anyons which has $45$ energy levels in the two-particle manifold, and, then, we derive the mean value $\braket{r}$ and inverse participation ratio (IPR) as functions of statistical angle $\theta$. The key idea is recording the response of the system after a local perturbation as a function of time and using spectrum analysis to reveal the characteristic modes of the system. We, here, consider only the two-particle energy manifold; although it is the simplest case for interacting systems, some typical features of the many-body localization emerge. The initial state is prepared in a product state,
\begin{equation}
\ket{\psi_0}_{m,n}=\cdots\left(\frac{\ket{0}+\ket{1}}{\sqrt{2}}\right)_m\cdots\left(\frac{\ket{0}+\ket{1}}{\sqrt{2}}\right)_n\cdots,
\end{equation}
where sites $m$ and $n$ are in superposition of $\ket{0}$ and $\ket{1}$, and all other sites are in $\ket{0}$ states. The state evolved at time $t$ reads
\begin{align}
\ket{\psi(t)}_{m,n}&=\frac{1}{2}\ket{\mathrm{Vac}}+\frac{1}{2}\sum_\beta C^\beta_{m,n}e^{-i[(E^{(2)}_\beta t)/\hbar]}\ket{\phi_\beta^{(2)}}\\
&+\frac{1}{2}\sum_\alpha (C^\alpha_{m}+C^\alpha_{n})e^{-i[(E^{(1)}_\alpha t)/\hbar]}\ket{\phi_\alpha^{(1)}},\nonumber
\end{align}
where $\ket{\phi^{(2)}_\beta}$ is the $\beta$-th eigenstate in the two-particle manifold with the corresponding energy $E^{(2)}_\beta$ and $C^\beta_{m,n}=\braket{\phi_\beta^{(2)}|1_m,1_n}$. $E^{(1)}_\alpha, \ket{\phi^{(1)}_\alpha}$, and $C^\alpha_{m}$ are counterparts in the single-particle manifold and are irrelevant in this simulation. The two-point correlation of a two-particle lowering operator can be expressed as
\begin{align}
\chi_2(m,n)&=\braket{(\sigma^x_m+i \sigma^y_m)(\sigma^x_n+i \sigma^y_n)}\\&=\braket{\sigma^x_m \sigma^x_n}-\braket{\sigma^y_m \sigma^y_n}+i\braket{\sigma^x_m \sigma^y_n}+i\braket{\sigma^y_m \sigma^x_n},\nonumber
\end{align}
where $\sigma^x=\ket{1}\bra{0}+\ket{0}\bra{1}$ and $\sigma^y=i\ket{1}\bra{0}-i\ket{0}\bra{1}$. The time dependent expectation value of the two-point correlation is
\begin{equation}
\chi^t_2(m,n)=\frac{1}{4}\sum_\beta|C^\beta_{m,n}|^2e^{-i[(E^{(2)}_\beta t)/\hbar]}.
\end{equation}
It is obvious that the single-particle component is projected out, and we can reveal $|C^\beta_{m,n}|^2$ and $E^{(2)}_\beta$ by the discrete-time Fourier transform,
\begin{equation}
\chi_2^f(m,n)[k]=\frac{1}{N}\sum_{l=1}^Ne^{-2\pi i(kl/N)}\chi_2^t(m,n)[l],
\end{equation}
where $N=T/\tau$ is the number of data sampled from evolution time $T$ with sampling interval $\tau$. By varying $m$ and $n$ in the initial state, the confidence of detecting each energy level is enhanced.

\begin{figure}[tb]
\centerline{\includegraphics[width=0.48\textwidth]{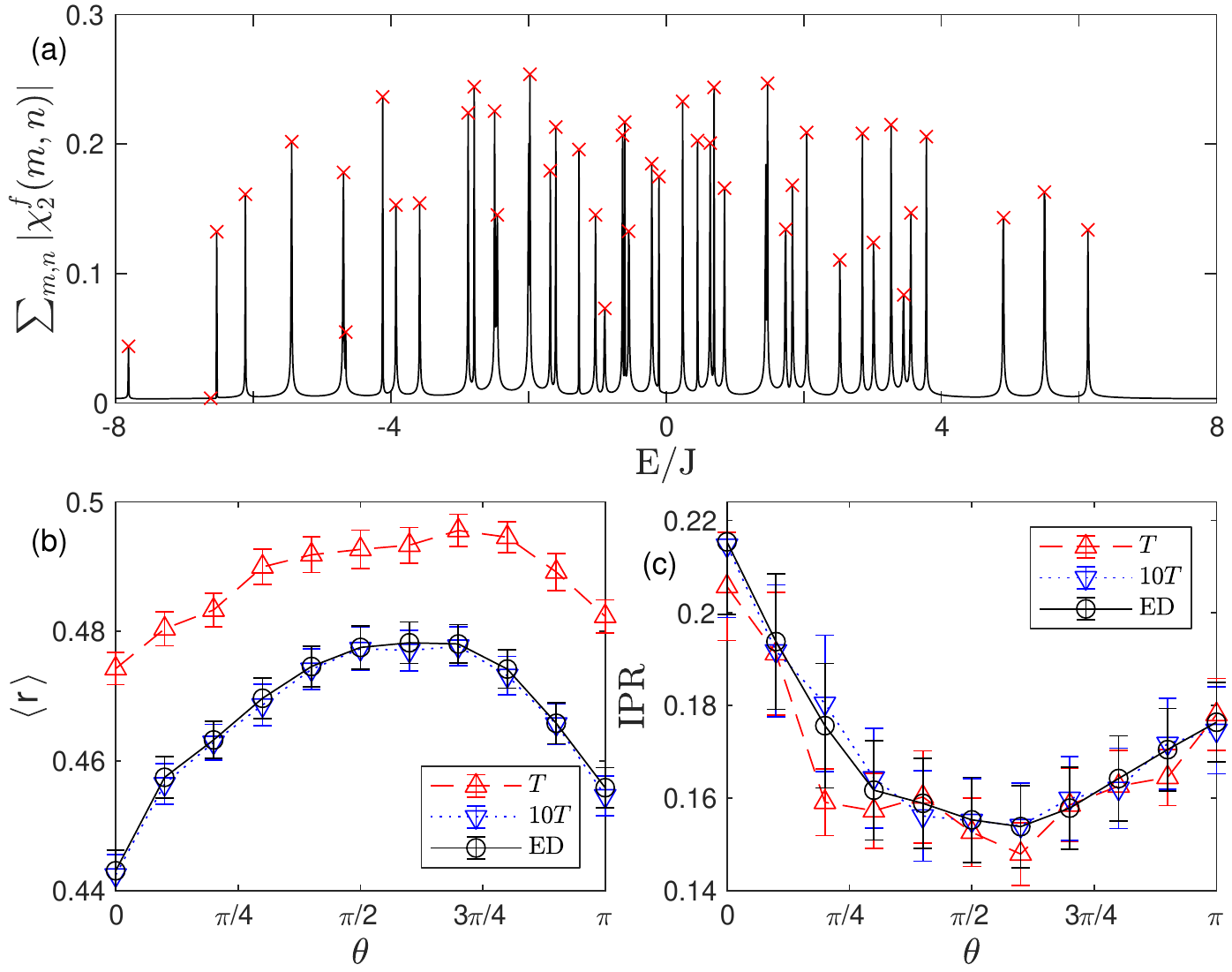}}%
 \caption{\label{fig:experiment}(Color online) (a) The discrete-time Fourier transform of the expectation value $\chi_2(m,n)$ as a function of energy (frequency) for a single disorder realization. Data are obtained from $2000$ time step samplings with $\theta=0.5\pi$ and a sum over every possible initial state. Red crosses are peaks corresponding to the eigenenergies of the disordered anyon-Hubbard model. (b) The mean value $\braket{r}$ as a function of statistical angle $\theta$. The black solid line is calculated from ED, the red dashed curve is data simulated with 2000 time step samplings, and the blue dotted line is simulated with 20000 samplings. (c) The IPR as a function of $\theta$. Other parameters are $J=1, U=1, W=3$, and $L=9$. (b) and (c) are averaged over $100$ disorder realizations.}
 \end{figure}

In Fig.~\ref{fig:experiment} (a), we plot $\sum_{m,n}|\chi_2^f(m,n)|$ as a function of energy (frequency) for a single disorder realization of the anyon-Hubbard model with the parameters $J=1, U=1, W=3$, and $\theta=0.5\pi$. The discrete-time Fourier transform  is preformed from data obtained by $2000$ time step samplings, and 44 eigenenergies can be resolved from the peaks labeled by red crosses. In Fig.~\ref{fig:experiment} (b), we present the mean value $\braket{r}$ as a function of $\theta$ calculated from the peak positions of the discrete-time Fourier transform of $N=2000$ samplings in evolution time $T$ (red upward triangle labeled), $N=20000$ samplings in evolution time $10T$ (blue downward triangle labeled), and ED results (black circle labeled), all data are averaged over $100$ disorder realizations. The non-monotonic relation $\braket{r (\theta)} $ can be clearly seen from the Fourier transform of $N=2000$ samplings. Here, the values of $\braket{r}$ are larger than ED results due to the energy level missing. One can measure after a longer evolution time and resolve the spectrum more accurately, but it requires much longer decoherence time of the experimental system.

Simultaneously, the IPR of each eigenstate in the representation of single-occupation states can be calculated as
\begin{equation}
\mathrm{IPR}_\beta=\frac{\sum_{m,n}|C^\beta_{m,n}|^4}{\left(\sum_{m,n}|C^\beta_{m,n}|^2\right)^2},
\end{equation}
where $|C^\beta_{m,n}|^2$ is accessible from the amplitude of the Fourier transform $|\chi_2^f(m,n)|$ at the corresponding energy (frequency) and the denominator is a normalization constant. We display the simulated IPR as a function of $\theta$ in Fig.~\ref{fig:experiment} (c) with the parameters $J=1, U=1$, and $W=3$. Here, $\beta$ is chosen at the middle of the energy spectrum, and the data are averaged over $100$ disorder realizations. The non monotonic relation is also distinctly shown by IPR calculated from $N=2000$ samplings (red upward triangle), $N=20000$ samplings (blue downward triangle), and the ED results (black circle). {\color{black} We note that the numerical simulation reveals the degree of localization properties rather than the localization-delocalization transition for systems with different $\theta$'s. In this sense, the systems with $\theta=0$ are more likely to be localized than others. In addition, our numerical simulation indicates the observation of the non monotonic $\theta$ dependence of the localization properties in a realistic experimental system of small size.}

\section{\label{sec:5}Discussion and conclusion}

{\color{black}Before concluding, we note that the non-monotonic behaviors are also observed for lower densities, such as the quarter-filling case. At low-density filling, the Pauli exclusion principle for quasi-fermions becomes distinct~\cite{WZhang2017}, and multi-occupancy states are suppressed. One may except that the localization properties of the one-dimensional anyon-Hubbard model would be independent of the statistical angle $\theta$ in the hard-core limit. However, in our numerical simulation, multi-\ particle occupancy is not ignorable for quarter-filling when $\theta=\pi$, and, thus, the on-site interaction is still relevant in this case. The density would play an import role in the quasi-momentum space in the clean limit of this model~\cite{Tang_2015} where the maximum momentum is non monotonically depending on $\theta$ and the non-monotonic behavior becomes more distinct when increasing density. It would be interesting to further investigate the dependence of the localization properties on the filling density.}

To summarize, we have explored the localization properties of the ergodic and localized phases in the one-dimensional disordered anyon-Hubbard model. Several physical characteristics, such as the half-chain entanglement, the adjacent energy-level gap-ratio parameter, the long-time limit of the particle imbalance, and the critical disorder strength, have been numerically calculated. It is found that these localization characters are non monotonically dependent on the anyon statistical angle. Furthermore, we have demonstrated that the statistics can induce localization-delocalization transition by studying the hybridization parameter. Finally, the possibility of observing these statistically related properties in the experiments is explored based on the numerical simulation of spectroscopy of energy levels.

\begin{acknowledgments}
This work was supported by the NKRDP of China (Grant No. 2016YFA0301800), the NSAF (Grants No. U1830111 and No. U1801661), the NSFC (Grant No. 11704132), the Key-Area Research and Development Program of Guangdong Province (Grant No. 2019B030330001), and the Key Program of Science and Technology of Guangzhou (Grant No. 201804020055).
\end{acknowledgments}

\bibliography{ref}

\end{document}